\documentstyle[12pt,epsf]{article}

\newcommand{\beq}{\begin{equation}}
\newcommand{\eeq}{\end{equation}}
\newcommand{\nn}{\nonumber}
\newcommand{\bea}{\begin{eqnarray}}
\newcommand{\eea}{\end{eqnarray}}

\newcommand{\gtrsim}{\ \rlap{\raise 2pt\hbox{$>$}}{\lower 2pt \hbox{$\sim$}}\ }
\newcommand{\lessim}{\ \rlap{\raise 2pt\hbox{$<$}}{\lower 2pt \hbox{$\sim$}}\ }

\newcommand{\np}[1]{Nucl. Phys. {\bf #1}}
\newcommand{\pl}[1]{Phys. Lett. {\bf #1}}
\newcommand{\pr}[1]{Phys. Rev. {\bf #1}}
\newcommand{\prl}[1]{Phys. Rev. Lett. {\bf #1}}
\newcommand{\zp}[1]{Z. Phys. {\bf #1}}

\newcommand{\arnps}[1]{{ Ann. Rev. Nucl. Part. Sci. }{\bf #1}}
\newcommand{\ptp}[1]{Prog. Theor. Phys. {\bf #1}}

\makeatletter
\if@twoside
\else \oddsidemargin 00pt \evensidemargin 00pt
\fi
\let\@eqnsel = \hfil

\expandafter\ifx\csname mathrm\endcsname\relax\def\mathrm#1{{\rm #1}}\fi
\@ifundefined{mathrm}{\def\mathrm#1{{\rm #1}}}{\relax}

\makeatother

\begin{document}

\title{\vskip-2.5truecm{\hfill \baselineskip 14pt {{
\small  \\
\hfill MZ-TH/98-40 \\ 
\hfill September  1998}}\vskip .9truecm}
 {\bf CP VIOLATING LEPTON ASYMMETRIES IN \\
LEFT-RIGHT MODELS}}

\vspace{5cm}

\author{Gabriela Barenboim\footnote{\tt gabriela@thep.physik.uni-mainz.de} 
\phantom{.}
 \\  \  \\
{\it  Institut f\H ur Physik - Theoretische 
Elementarteilchenphysik }\\
{\it Johannes Gutenberg-Universit\H at, D-55099 Mainz, Germany}
\\
}

\date{}
\maketitle
\vfill

\begin{abstract}
\baselineskip 20pt           

Lepton charge asymmetries can be used as an alternative means of
searching for new physics. They are interesting because they are
small in the Standard Model and therefore, necessarily evidence new
physics. 
In this work we explore the use of lepton asymmetries as a probe of
the flavour structure of the left-right symmetric model with
spontaneous CP violation. We find that new physics may enhance the
magnitude of $a_{\mbox{\tiny{SL}}}$  up to the precent level within the 
appropiate parameter space.

\end{abstract}
\vfill
\thispagestyle{empty}

\newpage
\pagestyle{plain}
\setcounter{page}{1}

\section{Introduction}

CP violation is one of the few remaining unresolved mysteries in
particle physics. Since its discovery, no general agreement has been
reached on its origin.
The simplest and most widely accepted explanation is due to Kobayashi and 
Maskawa. In their model CP is explicitly broken by the Yukawa couplings 
and no other fields than those of the minimal Glashow-Salam-Weinberg model
are required.

Although there is no conflict between the observation of CP violation
in the kaon system and this theory, intriguing hints of other plausible
explanations emerge from astrophysical consideration of the baryon to 
photon ratio in the Universe. On top of that, several extended models such
as supersymmetric models \cite{sup}, two Higgs doublet model
\cite{thd}, vector like quark models \cite{vlq}
and left-right symmetric models \cite{lrm,eg1,seis} 
(to mention only a few of them) can equally
well explain the existing data.
It is for that reason that exploration of CP violation in the $B$ system 
is so crucial.

The $B$ system offers several final states that provide a rich source 
for the study of this phenomenon. Even more, $B$ factories will provide us 
a unique tool to study CP violation in the $B$ system and, if we are
lucky enough will gift us with the first evidence for physics beyond
the Standard Model.
Whatever the new physics will turned out to be, its detection would
give information which will not be accessible to high energy colliders.
For example, if our future shows itself as left-right symmetric, then
a detailed study of the $B$ system will give us the opportunity to catch 
a glimpse on the right-handed scale.

Particularly important are the processes for which the Standard Model 
predicts vanishing or negligible small results.
In this case, both observation and non observation of such effects can 
be directly related to constrains on the physics which hides itself beyond
the Standard Model.
With this aim, the measurements of the dilepton charge asymmetry and 
the total lepton charge asymmetry may play a crucial role in
distinguishing between non standard models of new physics.

These lepton asymmetries measure the relative phase between 
$\Gamma_{12}$ and $M_{12}$. In the Standard Model, they have
(to leading order) the same phase and therefore these asymmetries are
strongly suppressed. In models of new physics, new contributions
to mass mixing which would generally carry a different
phase from that of the Standard Model, change completely the
landscape.
We will show that in the left-right symmetric model with spontaneous
CP violation one would expect more than an order of
magnitude enhancement over the Standard Model predictions for both
$B_d$ and $B_s$. We will also investigate the correlation between
the CP asymmetry $a_{\mbox{\tiny{SL}}}$ and the CP asymmetries in
$B_d \rightarrow J/\psi \; K_s$ and 
$B_d \rightarrow \pi^+ \; \pi^-$.

The remainder of this paper is organized as follows. We begin in section 
2 by presenting some necessary preliminarities to
$B^0$-$\overline{B^0}$  mixing. The basic features of the left-right 
symmetric model with spontaneous CP violation as well as the
new contributions of this model to $B^0$-$\overline{B^0}$  mixing
are given in section 3. Section 4 contains the results and the analysis of the 
above mentioned correlation. Finally, we conclude in section 5. 

\section{The formalism}
The general formalism to describe mixing among the neutral $B^0$ and
$\overline{B^0}$ mesons is completely analogous to the one in the kaon system.
Assuming CPT symmetry to hold, the $2\times 2$ $B^0$-$\overline{B^0}$ 
mixing matrix can be written as
\bea
{\cal{M}}=\pmatrix{ M & M_{12} \cr M_{12}^* &M} - \frac{i}{2}
\pmatrix{ \Gamma & \Gamma_{12} \cr \Gamma_{12}^* & \Gamma}
\eea
where the diagonal elements $M$ and $\Gamma$ are real parameters. If CP
were conserved, $M_{12}$ and $\Gamma_{12}$ would also be real.

So far everything looks similar to the kaon system, however, in the
$B$ system, the physical mass eigenstates
\bea
\mid B_{1,2} \rangle \; = \; \frac{1}{\sqrt{\mid p\mid^2 +
\mid q \mid^2}} \left( p \mid B^0 \rangle \, + \, 
q \mid \overline{B^0} \rangle \right) \; ,
\eea
where
\bea
\frac{q}{p}  = 
\left( \frac{M_{12}^* - \frac{i}{2} \Gamma_{12}^*}{
M_{12} - \frac{i}{2} \Gamma_{12}} \right)^{\frac{1}{2}} \; .
\label{qp}
\eea
whose masses are given by
\bea
M_{1,2}= M \pm {\mbox{Re}} \sqrt{ \left( M_{12} - \frac{i}{2}
 \Gamma_{12} \right)
\left( M_{12}^* - \frac{i}{2} \Gamma_{12}^*\right)} 
\eea
with
\bea
\Gamma_{1,2}= \Gamma \pm {\mbox{Im}} \sqrt{ \left( M_{12} - \frac{i}{2}
 \Gamma_{12} \right)
\left( M_{12}^* - \frac{i}{2} \Gamma_{12}^*\right)} 
\eea
have now  comparable lifetime, because many decay modes are common to both
states and therefore the available phase space is similar.
(Here we use the convention that $B^0$ contains a $b$ quark, thus
CP$\mid B^0 \rangle = - \mid \overline{B^0} \rangle$).

A measure of the mixing is given by $r$ and $\overline{r}$, the ratios
of the total (time-integrated) probabilities
\bea
r= \left| \frac{p}{q}\right| \frac{ \Delta M^2 + \left( \frac{1}{2} 
\Delta \Gamma \right)^2}{2 \Gamma^2 + \Delta M^2 - \left( \frac{1}{2} 
\Delta \Gamma \right)^2} \\
\nn\\
\overline{r}= \left| \frac{q}{p}\right| \frac{ \Delta M^2 + \left( \frac{1}{2} 
\Delta \Gamma \right)^2}{2 \Gamma^2 + \Delta M^2 - \left( \frac{1}{2} 
\Delta \Gamma \right)^2}
\eea
where $\Delta M = M_1 -M_2 $ and $\Delta \Gamma =\Gamma_1 -
\Gamma_2 $

The dilepton charge asymmetry, i.e. the asymmetry between the number of 
$l^+l^+$ and $l^-l^-$ pairs produced is defined as
\bea
a_{\mbox{\tiny{SL}}} \equiv  \frac{N(l^+l^+) - N(l^-l^-)}{N(l^+l^+) + N(l^-l^-)}
\eea
and it is easily found to be
\bea
a_{\mbox{\tiny{SL}}}=\frac{ \left| p/q \right|^2  - \left|q/p \right|^2}
{ \left| p/q \right|^2  + \left|q/p \right|^2} =
\frac{ {\mbox{Im}} \left( \Gamma_{12}/M_{12} \right)}
{1 + \frac{1}{4} \left| \Gamma_{12}/M_{12} \right| } \simeq
{\mbox{Im}} \left( \frac{\Gamma_{12}}{M_{12}} \right)
\label{cla}
\eea
This approximation holds if $\left|\Gamma_{12}/ M_{12} \right| \ll 1$ 
which is the 
case for the $B^0$-$\overline{B^0}$ system even in the presence of new
physics. In order for this relation to be violated, one needs a new dominant
contribution to the tree decays of $b$ mesons, which is extremely
unlikely or strong suppression of the mixing compared to the Standard Model
box diagram, which is also unlikely.

Unfortunately, this $\Delta B =2 $ asymmetry is expected to be quite tiny 
in the Standard Model, because $\left|\Delta \Gamma / \Delta M  \right| 
\simeq   \left|\Gamma_{12}/ M_{12} \right| \ll 1$ $\left( 
\Delta M \equiv M_{B_1} - M_{B_2} \, , \; \Delta \Gamma \equiv
\Gamma_1 -\Gamma_2 \right)$. 
This can be easily understood by looking to the relevant box diagrams
contributing to the $B^0$-$\overline{B^0}$ transition.
The mass mixing is dominated by the top-quark graphs, while the decay 
amplitude get obviously its main contribution from the $b \rightarrow c$
transition. Thus,
\bea
 \frac{\Gamma_{12}}{M_{12}} \simeq \frac{3 \pi}{2} \frac {m_b^2}{m_t^2}
\frac{1}{S(x_t)} \ll 1
\eea
where $S(x_t)$ \cite{ina} is a slowly increasing function of $x_t$
$\left( S(0)=1\,,\; S(\infty)=1/4 \right)$.
One has then (see Eq.(\ref{qp}))
\bea
\left| \frac{q}{p} \right| \simeq 1 + 
\frac{1}{2} \left| \frac{\Gamma_{12}}{M_{12}} \right| \sin 
\phi_B 
\eea 
where
\bea
\sin \phi_B \equiv {\mbox{arg}} \left( \frac{M_{12}}{\Gamma_{12}} \right)
\eea
The factor $\sin \phi_B$ involves and additional GIM suppression
\bea
\sin \phi_B \simeq \frac{8}{3} \frac{m_c^2 -m_u^2}{m_b^2}
{\mbox{Im}} \left(\frac{V_{cb} V_{cq}^*}{V_{tb} V_{tq}^*} \right)
\eea
implying a value of $\left| q/p \right|$ very close to 1. Here
$q\equiv d,s$ denote the corresponding CKM matrix elements for
$B_q^0$ mesons. Therefore, one expects,
\bea
a_{\mbox{\tiny{SL}}} \leq  \left\{ \begin{array}{ll}
10^{-3} \;  \; & \;\; (B_d^0) \\
10^{-4} \;\; & \;\; (B_s^0 )
\end{array}
\right.
\eea
within the Standard Model. Thus, a lepton asymmetry in excess of these values
is required in order to test new physics. Therefore, in this work this will
be our smoking-gun.
              
Another useful quantity to look at is the total lepton charge asymmetry,
which is defined by
\bea
l^\pm \equiv \frac{ N(l^+) - N(l^-)}{ N(l^+) + N(l^-)} = 
 \frac{N(l^+l^+) - N(l^-l^-)}{N(l^+l^+) + N(l^-l^-) + N(l^+ l^-) +
N(l^- l^+)}
\eea
This quantity is smaller than $a_{\mbox{\tiny{SL}}}$, 
but should be measured with better
statistics. Unlike Eq.(\ref{cla}) which is true whether or not 
$B^0$-$\overline{B^0}$ is produced coherently, the total lepton charge
asymmetry has a different form when expressed in terms of $r$ and
$\overline{r}$ in the case of a coherently or incoherently produced
$B^0$-$\overline{B^0}$ pair.
When it is produced coherently, the total lepton asymmetry is given by
\bea
l^\pm = \frac{r -\overline{r}}{2 + r + \overline{r}}
\sim 
\left\{ \begin{array}{ll}
.18 \; a_{\mbox{\tiny{SL}}} \;  \; & \;\; (B_d^0) \\
.43 \; a_{\mbox{\tiny{SL}}} \;\; & \;\; (B_s^0 )
\end{array}
\right.
\eea
where we have quoted the $B_s^0$ value for completeness as in the 
$B$-factories, just the $B^0_d$-$\overline{B^0}_d$ can be
coherently produced.
When the $B^0$-$\overline{B^0}$ is produced incoherently, the total lepton
asymmetry becomes
\bea
l^\pm = \frac{r -\overline{r}}{1 + r + \overline{r} + r \overline{r}}
\sim 
\left\{ \begin{array}{ll}
.29 \; a_{\mbox{\tiny{SL}}} \;  \; & \;\; (B_d^0) \\
.49 \;a_{\mbox{\tiny{SL}}}  \;\; & \;\; (B_s^0 )
\end{array}
\right.
\eea
The single lepton asymmetry measures the same quantity, Im$\left( \Gamma_{12}/
M_{12} \right)$, as the dilepton asymmetry. However, although as can be seen
the prediction for the dilepton asymmetry is bigger, as both leptons must be
tagged, the statistics is smaller.

\section{The model}

Left-right symmetric models \cite{uno} have been introduced more than twenty 
 years ago as a natural extension of the standard model, mainly in order
to justify on physical grounds the typical P-violating structure of 
weak interactions.

Starting from the gauge group $SU(2)_L \times SU(2)_R \times U(1) $ one is led
to assume an initial left-right symmetry of the Lagrangian, by ascribing
to the spontaneous symmetry breaking mechanism of the local gauge
symmetry the natural basis of the maximal parity violation at low energies.
As a consequence, parity restoration is to be expected beyond the mass
scale at which the spontaneous breakdown is supposed to happen.

According to the left-right symmetry requirements, quarks (and similarly
leptons) are symmetrically placed in left and right doublets
\bea
Q_{\alpha L}=\pmatrix{u_\alpha \cr d_\alpha }_L \equiv \left( 2,1,\frac{1}{3}
\right) \;,\;\;
Q_{\alpha R}=\pmatrix{u_\alpha \cr d_\alpha }_R \equiv \left( 2,1,\frac{1}{3}
\right) \nn
\eea
where $\alpha=1,2,3$ is the generation index, and the representation 
content with with respect to the gauge group is explicitly indicated.

Similarly, gauge vector bosons consist of two triplets 
{\bf{W}}$_L \equiv (3,1,0)$,
{\bf{W}}$_R  \equiv (1,3,0)$, and a singlet $B^\mu \equiv (1,1,1)$.
While the scalar content of the minimal model is given by a complex
bidoublet $\phi \equiv (2,2,0)$ and one set of left-right symmetric
lepton-number carrying triplets $\Delta_L \equiv (3,1,2)$ and
$\Delta_R \equiv (1,3,2)$.
(A model with only a minimal scalar sector and spontaneous CP violation
predicts unacceptably large FCNC \cite{fcnc}, and it is because of this that
one should add scalar triplets but these do not affect our analysis).
The vacuum expectation values of the fields have the following form,
\bea
\langle \phi \rangle = \pmatrix{k & 0 \cr 0 & k^{\prime} e^{i\alpha} }
\nn \\ \nn \\
\langle \Delta_L \rangle = \pmatrix{0 & 0 \cr 0 & v_L}
\\ \nn \\
\langle \Delta_R \rangle = \pmatrix{0 & 0 \cr 0 & v_R e^{i\theta}}\nn
\eea
The relative phases between $k$ and $k^{\prime}$, $\alpha$, and between
$v_L$ and $v_R$, $\theta$, spontaneously break CP. However the leptonic sector
does not concern us here, then, $\alpha$ will be the only source of CP
violation. Eventually, there are seven CP violating phases in the mass
eigenbasis, but they are all proportional to $\alpha$.

In the Standard Model, 
the dominant contribution to $B^0-\overline{B^0}$ mixing arises
from the top quark contribution to the box diagram. Let us assume that the off
diagonal element of $B^0_q-\overline{B^0}_q$ is changed by a factor
$\Delta_{qb}$ as a result of a new contribution from the left-right 
symmetric model
\bea
M_{12}= M_{12}^{(SM)} \Delta_{qb} \hspace{2cm} (q=d,s)
\label{par}
\eea
where $ M_{12}^{(SM)}$ is the 
standard (LL) box contribution which is given by \cite{ocho}
\bea
M_{12}^{SM} \simeq \frac{G_F^2}{12 \pi^2} B_B f_B^2 m_B M_1^2 
\eta_2^{(B)} S(x_t) (\lambda_t^{LL})^2 \; ,
\eea
where  $\eta_2^{(B)} \simeq 0.83 $ is a QCD correction factor,
$S(x_t)$ is a well known Inami-Lim function  and
\bea
\lambda_t^{AB} = V_{A,tl} V_{B,tb}^* \; \; \; \; \;(A,B=L,R)
\eea
while
\bea
\Delta_{qb} = 1 + \frac{M_{12}^{LR} }{ M_{12}^{SM} }.
\eea
The left-right  part gets its weight basically from three major sources, the
first one
is given by the $W_1 W_2$ box diagram
which is also dominated by the tt exchange \cite{siete}
\bea
M_{12}^{W_1 W_2} \simeq \frac{G_F^2}{\pi^2}  B_B M_1^2 \beta 
f_B^2 m_B \left( \left(\frac{m_B}{m_b + m_q} \right)^2 + \frac{1}{6}
\right) \lambda_t^{RL} \lambda_t^{LR} \eta_1^{LR}
F_1(x_t, x_b, \beta),
\eea 
the second one is the $S_1 W_2$ box ($S_1$ denotes the Higgs boson
that gives the longitudinal component to $W_1$)
\bea
M_{12}^{S_1 W_2} \simeq - \frac{G_F^2}{4 \pi^2} B_B M_1^2 \beta 
f_B^2 m_B \left( \left(\frac{m_B}{m_b + m_q} \right)^2 + \frac{1}{6}
\right) \lambda_t^{RL} \lambda_t^{LR} \eta_2^{LR}
F_2(x_t, x_b, \beta)
\eea 
with $\beta = M_1^2 /M_2^2$, $x_i = m_i^2 /M_1^2$ and  the QCD coefficients
$\eta_1^{LR} \approx \eta_2^{LR} \approx 1.8$.

$F_1$ is given by
\bea
F_1(x_t, x_b, \beta) = \frac{ x_t}{(1-x_t \beta)
(1-x_t)} \int_0^1 d\alpha \, {\sum_k}^\prime \ln \mid
\Lambda_k(\alpha,\beta)\mid
\eea
with
\bea
{\sum_k}^\prime  = \sum_{1,2} - \sum_{3,4}
\eea
and the functions $\Lambda_k(\alpha, \beta)$ are defined by
\bea
\Lambda_1(\alpha, \beta) &=&  x_t  - x_b \alpha (1 - \alpha) \; , \nn \\
\Lambda_2(\alpha, \beta) &=& 1 - \alpha + \frac{\alpha}{\beta} -
x_b \alpha (1 - \alpha) \; , \nn \\
\Lambda_3(\alpha, \beta) &=& x_t (1 - \alpha) + \frac{\alpha}{\beta}
- x_b \alpha (1 -\alpha) \; ,\nn \\
\Lambda_4(\alpha, \beta) &=& 1 - \alpha + x_t \alpha
- x_b \alpha (1 -\alpha) \; ,
\eea
and
\bea
F_2(x_t, x_b, \beta) = \frac{ 2 x_t }{(1-x_t \beta)
(1-x_t)} \int_0^1 d\alpha \, {\sum_k}^\prime  \Lambda_k(\alpha,\beta) 
\ln \mid\Lambda_k(\alpha,\beta) \mid \; .
\eea
The last contribution is given by the  tree level pieces 
which are mediated by the flavour changing
neutral Higgs bosons $\phi_1$ and $\phi_2$ and their contribution
is \cite{egn}
\bea
M_{12}^H \simeq - \frac{\sqrt{2} G_F}{M_H^2} m_t^2(m_b) B_B f_B^2 
m_B  \left( \left(\frac{m_B}{m_b + m_q} \right)^2 + \frac{1}{6}
\right) \lambda_t^{RL} \lambda_t^{LR} \; ,
\eea
where we have assumed a common Higgs mass $M_H$. 

Regarding $\Gamma_{12}$, as the tree level Higgs exchange does not
contribute and all the terms containing a $W_R$ have a suppressing $\beta$
factor, the left-right contribution is completely negligible and
we can take that $\Gamma_{12} \simeq \Gamma_{12}^{SM}$.

Consequently, it is clear that, because of the new contributions
to $M_{12}$, the relative phase between $\Gamma_{12}$ and $M_{12}$ can 
now, unlike in the Standard Model case,  be different.
In fact, we have now that $\Delta_{bd}$, which parameterizes the
deviation from the Standard Model box diagram can be written as
\bea
\Delta_{bq}= 1 + \rho e^{i \sigma}
\eea
where
\bea
\rho &\equiv & \left| \frac{M_{12}^{LR}}{M_{12}^{SM}} \right| \nn\\
&=& 
\frac{\beta \left(\left( \frac{m_B}{m_b +m_q}\right)^2 + \frac{1}{6}\right) 
}{12 S(x_t) \eta_2^{(B)} }\left( F_1(x_t,x_b,\beta) \eta_1^{LR} -
\frac{1}{4} \eta_2^{LR}
 F_2(x_t,x_b,\beta) - \frac{\sqrt{2} m_t^2}{M_H^2 M_1^2 G_F}\right)
 \\
&\simeq & \left[ .21 + .13 \; \log\left(\frac{M_2}{1.6 {\mbox{TeV}}}\right)
\right] \left(\frac{1.6 {\mbox{TeV}}}{M_2}\right)^2 +
\left(\frac{12 {\mbox{TeV}}}{M_H}\right)^2 \nn
\eea
and
\bea
e^ { i \sigma} =  \frac{\lambda_t^{LR}\lambda_t^{RL} }{(\lambda_t^{LL})^2 }
= - \frac{V_{R,tq}V_{R,tb}^* }{V_{L,tq}V_{L,tb}^*  } 
\eea
with \cite{eg1}
\bea
\sin\sigma^{(d)} &\simeq & \eta_d \eta_b r
 \sin\alpha \left[
\frac{2 \mu_c}{\mu_s} \left( 1 + \frac{s_1^2 \mu_s}{2 \mu_d} \right)
+ \frac{\mu_t}{\mu_b} \right]
\; ,\nn \\
\sin\sigma^{(s)} &\simeq &\eta_s \eta_b r
 \sin\alpha \left[ \frac{\mu_c}
{\mu_s} + \frac{\mu_t}{\mu_b} \right] \; ,
\label{ang}
\eea
to first order in $ r  \sin\alpha$ and
\bea
 r=\frac{k}{k^\prime} \;\; \;, \; \;\;   \mu_i = \eta_i m_i \;\; \;, \; \;\; \eta_i^2=1 \;.
\eea
A detailed discussion of these formulae can be found in \cite{bs}.
The important 
point to notice is that the predictivity of this model resides in the fact that
by imposing spontaneous breakdown of CP, all the phases appearing in both
$V_L$ and $V_R$ depend on $\alpha$, as pointed out before. Besides that, 
due to the enhancement factor $m_t/m_b$ in Eq. (\ref{ang}) it is easy to see
that $\sigma^{(s),(d)}$ is by no means a suppression factor as can be 
naturally of order one.
In fact, it has a lower bound (coming from the 
$K$ system\cite{seis}) that ensures $\mid \sin\sigma^{(s),(d)} \mid \ge .2 $
 which guarantees that if nature is left right symmetric, this 
indeed will show up in CP asymmetries.

\section{Results}

We can now readily obtain an expression for the dilepton charge
asymmetry in the left-right model,
\bea
a_{\mbox{\tiny{SL}}}=
{\mbox{Im}}\left(\frac{\Gamma_{12}^{SM}}{M_{12}^{SM} + M_{12}^{LR}}
\right) \simeq \left(\frac{\Gamma_{12}}{M_{12}}
\right)_{SM} \left(- \frac{\rho \sin\sigma }{1+2\rho \cos\sigma + \rho^2}
\right)
\label{as}
\eea
Of course, when the denominator of Eq.(\ref{as}) vanishes, there 
exist other small contributions which eliminate the singularity.
These contributions are proportional to $\beta^2$ and arise from
the $W_2 - W_2$ and $W_1 - S_2$ box diagrams ($S_2$ denotes the
Higgs boson that gives the longitudinal component to $W_2$).
 
The result in Eq.(\ref{as}) shows explicitly that the dilepton charge
asymmetry can be strongly enhanced over the Standard Model value.
It is important to notice that such enhancement, which is larger when $\sigma$
is larger, especially when $\sigma$ is close to $\pi$, reaches its maximum
when $\rho=1$, for which it gives $1/2 \tan(\sigma/2)$.
But values of $\rho $ close to one are precisely the ones one expects in order
to have a left-right symmetric model with low energy observable
implications.

In Fig.1 we plot the value of $\rho $ as a function of $M_2$ fixing the Higgs
boson mass to 12 TeV and in Fig.2 as a function of $M_H$ fixing the 
right-handed gauge boson mass to 1.6 TeV.
From the figures, it can be easily seen that for Higgs and gauge boson masses
close to their current lower bounds the enhancement is maximal.

We present the absolute value of the
enhancement factor in Fig.3 for different values of
$\rho $ as a function of the spontaneous symmetry breaking phase $\alpha$.
With the chosen values for the masses, we conclude that left-right 
symmetry can enhance the dilepton charge asymmetry (and hence 
also the total lepton charge asymmetry) for more than one order of
magnitude.

At this point, it is important to notice that although in this work
we are focusing ourselves on the lepton charge asymmetry as a means
of searching for new physics, i.e. new physics in the mixing 
on the $B$-system to which these lepton asymmetries are sensitive, 
these new sources of mixing can be independently tested in other 
measurements. Particularly interesting alternative test for CP
violation in the $B$-system are the study of the CP asymmetries in
the $B_d \rightarrow J/\psi \; K_s$ and 
$B_d \rightarrow \pi^+ \; \pi^-$  decays, where new physics effects might 
be large and  have been
calculated by a large number of groups \cite{bur}.

Even though these studies offer promising possibilities it will be
very useful to have alternative means for searching new physics.
Precisely because the rate in the Standard Model is negligible, the
lepton asymmetries would be an extremely important measurement.

It should be emphasized that as the CP asymmetries in the decays
 $B_d \rightarrow J/\psi \; K_s$ and 
$B_d \rightarrow \pi^+ \; \pi^-$ are predicted (with a rather 
poor accuracy) to be sizeable in the Standard Model \cite{bf}
(the CP
asymmetries in $B_s^0$ decays which are supposed to be negligible
small in the Standard Model cannot be studied in the $B$ 
factories running at the $\Upsilon $ peak) a clear detection of
new physics by their measurement alone  is  an extremely
difficult task. 

In the left-right symmetric model with spontaneous
CP violation we are considering here, the $B_d^0$ CP asymmetries 
are given by
\bea
a _{J/\psi \, K_s } = \sin \left( 2 \beta + 2 \phi \right) \nn \\
a_{\pi \pi}=\sin \left( 2 \alpha + 2 \phi \right) 
\eea
where $\alpha$ and $\beta$ are  defined as usual
\bea
\alpha &\equiv& \mbox{arg} \left(- \frac{V_{L,td} V_{L,tb}^*}
{V_{L,ud} V_{L,ub}^*}
\right) \nn \\
&&\\
\beta &\equiv& 
\mbox{arg} \left( -  \frac{V_{L,cd} V_{L,cb}^*}{V_{L,td} V_{L,tb}^*}
\right) \nn
\eea
and
\bea
\sin \phi = \frac { \rho \cos\sigma^{(d)}}{1+ 2 \rho \cos\sigma^{(d)} + 
\rho^2}
\label{tro}
\eea
As can be seen from the above formulae, even if in this case the
new physics effect is big (notice however that the enhancement factor
here is smaller than in $a_{\mbox{\tiny{SL}}}$) 
it cannot change drastically the value of the
CP asymmetries because they are already expected to be of order one within
the Standard Model. 

Nevertheless new physics coming from the left-right model 
can change the sign
of the CP asymmetries and this indeed will be a clear signal of
physics beyond the Standard Model.  In any case, the measurement of both,
the CP asymmetries in $B_d^0$ decays and $a_{\mbox{\tiny{SL}}}$
can prove (or disprove) the left-right symmetric model due to the
existence of a strong correlation between them, as it is evident from
Eq.(\ref{as}) and Eq.(\ref{tro}).

\section{Conclusions}

In summary, it would be very interesting to perform an accurate 
measurement of the single lepton and/or the dilepton asymmetries.
The presence of new physics in $B^0_d - \overline{B}^0_d$ mixing can
enhance the magnitude of $a_{\mbox{\tiny{SL}}}$ to an observable level.
For instance, $\mid a_{\mbox{\tiny{SL}}}\mid \sim  10^{-2}$ can be
achieved within the left-right symmetric model with spontaneous CP
violation when the right-handed gauge boson and the flavour
changing Higgs boson masses are close to their present lower bound.
The reason is that while the relative phase between the Standard
Model quantities $M_{12}^{SM}$ and $\Gamma_{12}$ is negligible, the
relative phase between any of them and 
the left-right contribution $M_{12}^{LR}$ can be
quite large.

If the CP asymmetry $a_{\mbox{\tiny{SL}}}$ is really of the order of
$10^{-2}$ it should be detected at the forthcoming $B$ meson
factories, where as many as $10^8$ $B^0_d - \overline{B}^0_d$
events per year will be produced at the $\Upsilon$ resonance.
In fact, the experimental sensitivities to $a_{\mbox{\tiny{SL}}}$
are expected to be well within few percent
for both single lepton or dilepton asymmetry measurements \cite{yama},
 provided the number of  $B^0_d - \overline{B}^0_d$ exceeds $10^7$.
  
Of course the lepton asymmetries are correlated with the CP
asymmetries in the  $B_d \rightarrow J/\psi \; K_s$ and 
$B_d \rightarrow \pi^+ \; \pi^-$ processes. Measuring such a correlation
may partly testify or rule out the left-right symmetric model
with spontaneous CP violation.

\vspace{.5cm}

\begin{center}
{\bf Acknowledgements}
\end{center}
We are grateful to J.Bernabeu and M.Raidal for useful conversations.
A post-doctoral fellowship
of the Graduiertenkolleg `` Elementarteilchenphysik bei 
mittleren und hohen Energien"
of the University of Mainz is also acknowledged.

\clearpage

\begin{figure}
\begin{center}
\epsfxsize = 13cm
\epsffile{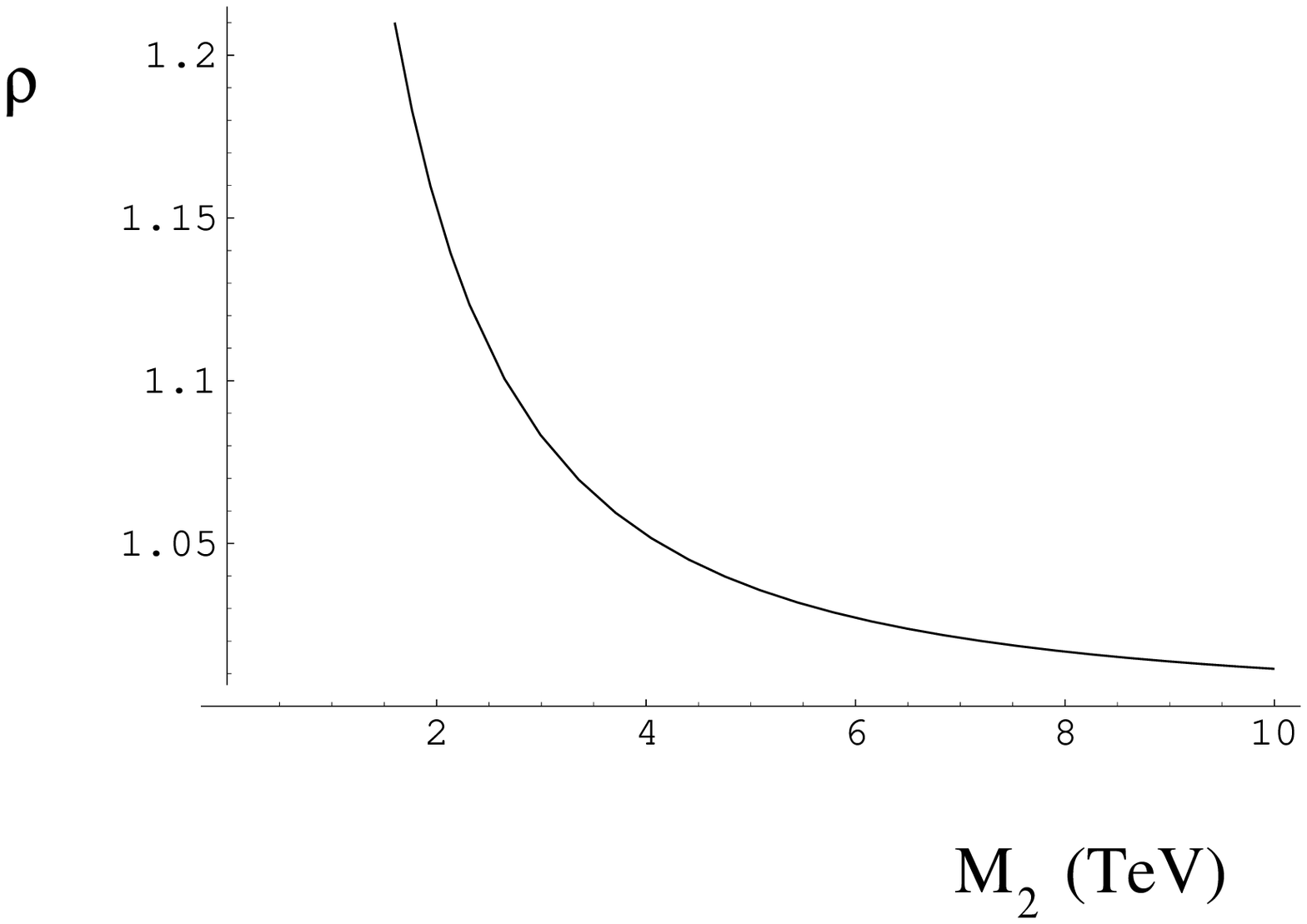}
\leavevmode
\end{center}
\caption{$\rho$ as a function of the right-handed gauge boson mass
for a fixed $M_H=12$ TeV. }
\end{figure}

\pagebreak
\begin{figure}
\begin{center}
\epsfxsize = 13cm
\epsffile{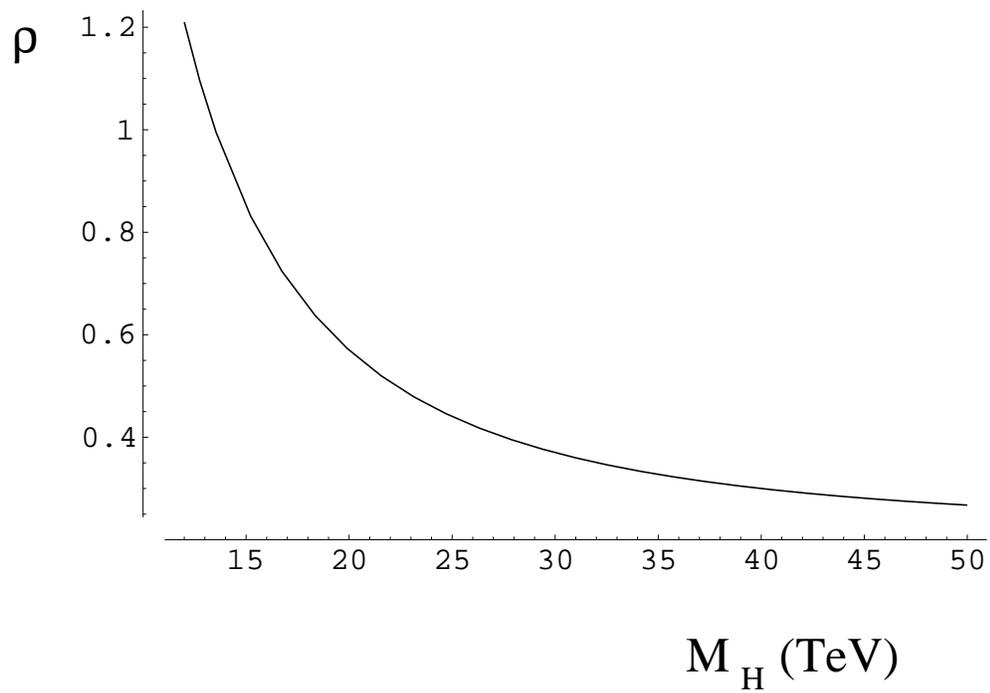}
\leavevmode
\end{center}
\caption{$\rho$ as a function of the flavour changing Higgs
boson massfor a fixed $M_2=1.6$ TeV.}
\end{figure}

\clearpage

\begin{figure}
\begin{center}
\epsfxsize = 13cm
\epsffile{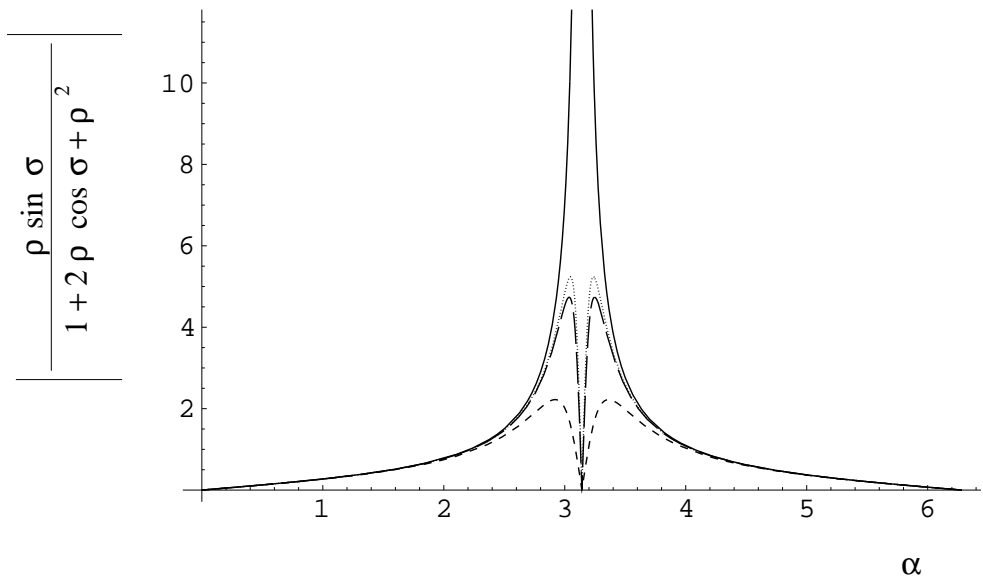}
\leavevmode
\end{center}
\caption{Absolute value of the enhancement factor, 
$ \frac{ \rho \sin \sigma }{ 1+2\rho \cos  \sigma + \rho^2} $ 
as a function of the spontaneous
symmetry breaking phase $\alpha $ for $\rho$=.8 (small dashed line), .9
(big dashed line), 1 (solid line), 1.1 (dotted line). }
\end{figure}

\end{document}